\documentclass[%
 reprint,
nofootinbib,
 amsmath,amssymb,
 aps,
]{revtex4-1}

\usepackage{graphicx}% Include figure files
\usepackage{dcolumn}% Align table columns on decimal point
\usepackage{bm}% bold math
\begin{document}

\preprint{APS/123-QED}

\title{Fuzzballs and black hole thermodynamics}% Force line breaks with \\

\author{Samir D. Mathur}
 %\altaffiliation[Also at ]{Department of Physics,\\ The Ohio State University,\\ Columbus,
%OH 43210, USA}%Lines break automatically or can be forced with \\
 \email{mathur.16@osu.edu}
\affiliation{Department of Physics, The Ohio State University, Columbus,
OH 43210, USA
}%


\def\nn{\nonumber \\}
\def\p{\partial}
\def\t{\tilde}
\def\h{{1\over 2}}
\def\be{\begin{equation}}
\def\bea{\begin{eqnarray}}
\def\ee{\end{equation}}
\def\eea{\end{eqnarray}}
\def\b{\bigskip}


\begin{abstract}

The fuzzball construction resolves the black hole information paradox by making spacetime end just before the horizon is reached.  But if there is no traditional horizon, then do we lose the elegant relations of black hole thermodynamics? Using an argument similar to modular invariance, we argue that the answer is no; the completeness of fuzzball states implies that the generic fuzzball indeed reproduces the thermal properties  attributed to the traditional hole.

\end{abstract}

%Use showkeys class option if keyword
\maketitle

%\tableofcontents

\section{\label{secone}Introduction}

In 3+1 dimensional Einstein gravity, black hole solutions are uniquely determined by their conserved quantum numbers (i.e., `black holes have no hair'). The region around the horizon is in the vacuum state for all quantum fields. Quantum mechanics in this black hole background exhibits an intriguing thermodynamics: the hole has an entropy $S$, a temperature $T$ and an energy $E$ satisfying $dE=TdS$. Further, the hole radiates quanta in a thermal distribution with temperature $T$ \cite{bek,hawking}.

But the mechanism of radiation involves pair creation at the horizon, and the radiation gets progressively more entangled with the remaining hole. This situation creates a problem near the endpoint of evaporation, where the small mass `remnant' must have an enormous number of  internal states to carry the required entanglement \cite{hawking2}.

\begin{figure}[h]
\includegraphics[scale=.32]{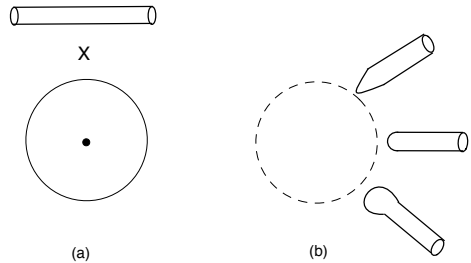}% Here is how to import EPS art
\caption{\label{c6} (a) In the traditional hole, compact directions appear as a tensor product with the noncompact ones (the tensor product is denoted by the cross). (b) In a fuzzball microstate the compact directions `cap-off' before the horizon is reached; different choices of  `caps' at the various angular positions  yield the entropy $S_{fuzzballls}$.}
\end{figure}

String theory does not appear to have such high degeneracy remnants, so one needs to find some mechanism that prevents the monotonic growth of entanglement during the radiation process. In \cite{cern} it was proved, using strong subadditivity, that small corrections cannot achieve this goal; we need a correction of order unity to the dynamics at the horizon.\footnote{See also \cite{giddings,avery,acp}.} In string theory we find the fuzzball construction \cite{fuzzballs}, where a nonperturbative effect eludes the no-hair theorems \cite{gibbonswarner} and provides the required alteration of the hole.  In a fuzzball microstate the spacetime ends just outside the horizon, because compact directions `cap-off' (fig.\ref{c6}). The structure of the cap is   supported by the fluxes, branes etc. present in the theory. The fuzzball proposal then says that all microstates of the hole are fuzzballs - i.e., no microstate has a traditional horizon. This alteration of the horizon removes the information paradox; the fuzzball radiates from its surface just like a piece of coal. 

While this proposal resolves the information issue, one might ask a second question: if the horizon is not the traditional one, then do we still get black hole thermodynamics with the traditionally computed values of $S, T$, and the Hawking radiation rate $\Gamma$?\footnote{In \cite{eternal} the assumption of standard black hole thermodynamics was used to argue for a holographic dual of the eternal hole.} For example neutron stars are also gravitating objects without a horizon, but their entropy and temperature are not given by the spacetime geometry of the star. For fuzzballs on the other hand the situation seems to be different. Radiation has been computed from a very simple set of fuzzballs \cite{ross}, and the rate of radiation was found to {\it exactly} agree with the Hawking emission rate predicted for these special fuzzballs \cite{cm1}.

In this paper we begin by noting that all thermodynamic properties of fuzzballs will agree with the traditional hole if {\it one} of them agrees; say the entropy $S$. We will then recall the Gibbons-Hawking computation of $S_{bh}$ for traditional black holes \cite{gibbons,hawkingisrael}, and see how this derivation needs to be modified for the situation where microstates are fuzzballs. This argument will give us $S_{bh}=S_{fuzzball}$.

\section{\label{sectwo}Relation between thermodynamic quantities}

 The entropy of fuzzballs is given by
  \be
 S_{fuzzball}(M)\equiv \ln {\cal N}_{fuzzball}(M)
 \ee
 where $N_{fuzzball}(M)$ is the number of fuzzballs  at mass $M$.
 
 Suppose we knew that the entropy of fuzzballs at  mass $M$ agreed with the  traditional black hole entropy at mass $M$
 \be
 S_{fuzzball}(M)=
 S_{bh}(M)
 \label{twtwo}
 \ee
 Then the temperature of the typical fuzzball will agree with the traditional black hole temperature 
\be
T_{fuzzball}(M)=T_{bh}(M)
\label{twthree}
\ee
since in both cases we have (setting $E=M$)
\be
TdS=dE
\ee
(For black holes this is the first law of black hole mechanics.  Fuzzball states describe a normal statistical system with a large number of states, so we again have the usual first law.)

The radiation rates will agree as well, for the following reason. For both black holes and normal radiating bodies, the radiation rate $\Gamma$ is connected to the absorption cross section $\sigma$ by
\be
\Gamma(\omega)={\sigma(\omega)\over e^{\omega\over T}-1}
\label{twone}
\ee
Consider first the computation of $\sigma(\omega)$ for the traditional black hole. This computation starts with an incoming wave at infinity, which is partially reflected back  from the geometry in the region $r-2M\sim M$. Once the wave reaches $r-2M\ll M$, it does not reflect any further; it continues to fall to the horizon $r=2M$ where it is completely absorbed. Thus in computing $\sigma(\omega)$ we find  the part of the incoming wave that makes it to the region $r-2M\ll M$ without being backscattered.

Now consider the computation of $\sigma(\omega)$ for absorption into a fuzzball. The fuzzball surface is presumably a few planck lengths outside $r=2M$, so the metric in the region $r-2M\sim M$ is the same as the metric of the traditional hole. Further, the part of the wave that reaches the fuzzball surface is absorbed almost completely, because of the large number of degrees of freedom of the fuzzball. Thus we will have $\sigma_{fuzzball}\rightarrow \sigma_{bh}$ in the limit of large black holes $M/m_p\rightarrow\infty$.\footnote{I thank Borun Chowdhury for pointing out this agreement of cross sections.}  

From (\ref{twone}) we  see that if $T_{fuzzball}=T_{bh}$, then we will have
\be
\Gamma_{fuzzball}=\Gamma_{bh}
\label{twfour}
\ee
To summarize, if we are given the agreement of entropies (\ref{twtwo}), then the agreement of temperatures (\ref{twthree}) and radiation rates (\ref{twfour}) follow. 

%\section{A rough analogy explaining the idea of the argument}

Let us now ask why we would expect $S_{fuzzball}=S_{bh}$. To see the nature of the argument that we will make, consider the example of a 2-d CFT with central charge $c$. The number of states at energy level $N$, in the limit of large $N$, is \cite{cardy}
\be
{\cal N}\sim e^{2\pi \sqrt{cN\over 6}}
\label{thir}
\ee
For a general interacting CFT, it would be very hard to prove this relation by constructing all the states at level $N$ and counting them. But the answer can be obtained quickly from modular invariance. For a theory that comes from a local Lagrangian, we can evaluate the partition function $Z$ on a torus through Hamiltonian evolution along the two different cycles  $\sigma, \tau$ 
\be
Z={\rm tr}\,  e^{-H_\sigma \Delta \sigma}={\rm tr}\,  e^{-H_\tau \Delta \tau}
\ee
Here $\Delta \sigma, \Delta \tau$ are the lengths of the torus cycles in the $\sigma, \tau$ directions. Taking ${\Delta \sigma\over \Delta \tau}\rightarrow\infty$ one can derive the relation (\ref{thir}). 

Similarly, it would be very hard to compute the number of fuzzballs ${\cal N}_{fuzzball}$ for a black hole of mass $M$ by constructing all the relevant fuzzballs and counting them. But if the theory arises from a local Lagrangian, then there can be an alternative way of getting the count of states.

 We first review the computation given in \cite{hawkingisrael} of the entropy for the traditional hole, and then see how a somewhat modified argument arises in a theory where the microstates are fuzzballs. 

\section{\label{secthree}The Gibbons-Hawking argument for entropy}

\begin{figure}[h]
\includegraphics[scale=.42]{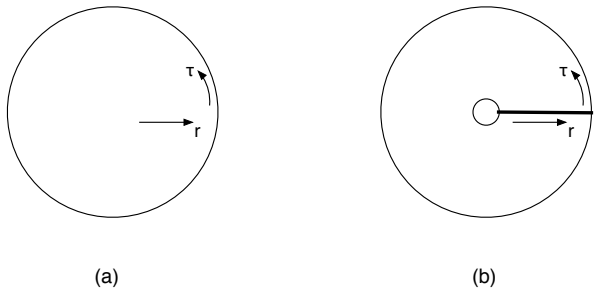}% Here is how to import EPS art
\caption{\label{c1} (a) The $r, \tau$ space is a `cigar' with the topology of a disc. (b) The bold line is the spatial slice on which a state of energy $E=M$ lives. Euclidean time evolution is in the $\tau$ direction. A small hole is cut around the point $r=2M$ where slices of different $\tau$ meet.}
\end{figure}

If we perform a Euclidean continuation $t\rightarrow -i\tau$, the Schwarzschild metric becomes
\be
ds_E^2=(1-{2GM\over r})d\tau^2+(1-{2GM\over r})^{-1}dr^2+r^2d\Omega_2^2
\label{fourt}
\ee
We compactify the $\tau$ circle with period $8\pi GM$; this makes (\ref{fourt}) have the shape of a `cigar' which ends smoothly at $r=2M$. The $r, \tau$ space is shown in fig.\ref{c1}(a). 

Let us now recall the steps given in \cite{hawkingisrael} for computing the black hole entropy. 

(a) The Euclidean action for gravity is
\be
S={1\over 16\pi G}\int R\sqrt{g} d^4 x +{1\over 8\pi G}\int_\p K \sqrt{h} d^3 x
\ee
where the second term is the boundary action. For the configuration (\ref{fourt}), the bulk term vanishes since the curvature scalar $R$ is zero. The boundary contribution diverges, but we subtract the value for flat space with $\tau$ again compactified  with period $8\pi GM$. This gives 
\be
I_M-I_{flat}=4\pi GM^2
\label{thone}
\ee
which yields
\be
Z=e^{-4\pi GM^2}
\label{sevent}
\ee

(b) Now we try to compute the same $Z$ in a Hamiltonian formulation, where the time evolution is in the $\tau$ direction. The spatial slices stretch in the $r$ direction from $r=2M$ to $r=\infty$. But these slices all meet at $r=2M$ (the tip of the `cigar'), so this is not a situation where we have the topology  ``spatial slice $\times$  time circle" required for Hamiltonian evolution. To remedy this problem, we cut a small disc out of $r, \tau$ plane around the point $r=2M$; the  resulting spatial slices are depicted in fig.\ref{c1}(b). We  will add back the contribution of the cut-out disc as a separate term.  We thus have  two contributions:

\b

(i) Writing (\ref{thone}) as 
\be
\log Z=-{\beta^2\over 16\pi G}
\ee
We have
\be
E=-{\p\over \p\beta} \log Z={\beta\over 8\pi G}=M
\label{thtwo}
\ee
 Thus the path integral with the $\tau$ circle identified with period $\beta=8\pi GM$ is peaked at configurations with energy  $E=M$. We take this energy  on the spatial slice of fig.\ref{c1}(b), and note that this slice evolves for a time $\Delta \tau=8\pi GM$. This gives a contribution to the partition function $Z$ equal to
\be
e^{-E\Delta\tau}=e^{-8\pi GM^2}
\label{fift}
\ee

(ii) The contribution of the small disc. The bulk action gives zero since the curvature scalar $R$ is zero, while the boundary gives $4\pi G M^2$. Thus we get a contribution to $Z$ equalling
\be
e^{4\pi G M^2}
\label{sixt}
\ee
Putting together (\ref{fift}) and (\ref{sixt}), we get
\be
Z=e^{-4\pi GM^2}
\ee
in agreement with (\ref{sevent}). Thus the Hamiltonian evolution method agrees with the path integral method.

Finally, we find the interpretation for the contribution of the small disc (\ref{sixt}). We have $
F\equiv -T\ln Z =E-TS $. This gives
\be
\ln Z=- {E\over T}+S
\label{twenty}
\ee
The first term on the RHS is the contribution (\ref{fift}). Thus we find that the entropy $S$ is given by the contribution (\ref{sixt}) 
\be
S_{bh}=4\pi G M^2
\ee
which can be seen to equal the well known expression ${A\over 4G}$.

\b

There are some issues of concern in this derivation:

\b

 (i) Once we cut out the small disc around $r=2M$, the state in (\ref{fift}) is defined on a spatial slice that has an inner boundary. What is the significance of a state having such a boundary? Is it clear that such a  state should be assigned the energy $M$?

\b

 (ii) While $S$ behaves like the entropy in the thermodynamic relation (\ref{twenty}), it is not clear why the entropy arrived at this way should be related in any way to a count of microstates. In particular, we do not see the differences between the $Exp[4\pi GM^2]$ orthogonal states that are predicted by this value of the entropy. 

\section{\label{secfive}Modifications for the case where microstates are  fuzzballs}

Now let us assume that we have a theory of gravity where all the microstates are fuzzballs. Our question is: what will be the number ${\cal N}$ of these fuzzballs?

\begin{figure}[b]
\includegraphics[scale=.32]{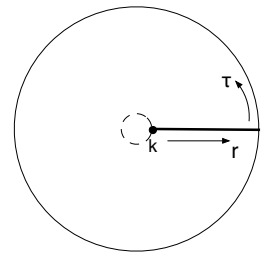}% Here is how to import EPS art
\caption{\label{c2} A fuzzball microstate (defined on slice depicted by the bold line) `caps-off'  before reaching the horizon; the cap is denoted by the black dot. Thus we do not need to cut out a hole around $r=2M$, but do have to take into account the degeneracy factor ${\cal N}_{fuzzball}$ arising from different possible caps. }
\end{figure}

We will again follow a path similar to the one above where we compared the path integral computation of $Z$ to a Hamiltonian computation of $Z$. We first outline the steps, and in the following section comment on the physical picture underlying the computation. 

\b

(a) We  work with the Euclidean theory and compactify the time direction $\tau$ with period $8\pi GM$, just as before.  We consider the path integral over all metrics that have mass $M$ at infinity. We assume that the solution (\ref{fourt}) is a saddle point to this path integral, thus getting 
\be
Z=e^{-4\pi GM^2}
\label{seventq}
\ee
 (This  step is the same as  the corresponding step in the Gibbons-Hawking argument.)  

(b) Now consider the Hamiltonian evolution along the $\tau$ direction. In Hawking's argument, the spatial slices all met at $r=2M$, so we had to cut out a small disc around this location. But fuzzball microstates `cap-off' before the horizon location $r=2M$ is reached (fig.\ref{c2}). Thus there is no disc that we need to cut out, and consequently, no contribution from the action of such a disc. Instead, we have the following contributions:

\b

(i)  By the same reasoning as in (\ref{thtwo}), the states contributing to the path integral are peaked around the energy $E=M$. Any one such state contributes to $Z$ an amount
\be
e^{-E\Delta\tau}=e^{-8\pi GM^2}
\label{fiftq}
\ee
(This is the same as (\ref{fift}.)

\b

(ii) We have a degeneracy factor from the region near $r=2M$  of the spatial slice, where we get different `caps' for different microstates. Letting ${\cal N}_{fuzzball}$ be the number of fuzzballs of mass $M$, we get\footnote{Fuzzball solutions have been constructed in a variety of ways; sometimes as coherent states (described by their classical mean value) and sometimes as energy eigenstates. Here we assume that we are using a basis made of energy eigenstates which we call $|F_a\rangle$.} 
\be
Z={\cal N}_{fuzzball}\, e^{-8\pi G M^2}
\label{eighttq}
\ee

Now equating (\ref{seventq}) to (\ref{eighttq}) we get
\be
{\cal N}_{fuzzball}=e^{4\pi G M^2}
\ee
This gives the entropy
\be
S_{fuzzball}\equiv \ln {\cal N}_{fuzzball}=4\pi G M^2={A\over 4G}
\label{twfive}
\ee
This is the same as the entropy computed in the Hawking computation; i.e., we get the desired relation (\ref{twtwo}). The relation (\ref{twfive}) is the main result of this paper. A key point in the argument is the fact that fuzzball states are wavefunctionals defined on the space of  manifolds that have no boundary near $r=2M$; instead these manifolds have a topology that allows them to `cap-off'. Because of this `capping-off',  the spatial slices do not meet at a common point $r=2M$ the way they did in the computation of \cite{hawkingisrael}. Consequently,  we do not need to cut out a disc around $r=2M$ to  define a good Hamiltonian evolution. Thus we do not get a contribution to $Z$ from the gravitational action of the disc, but in its place we get something more physical: an explicit degeneracy factor ${\cal N}_{fuzzball}$ from the different possible `cap' states.

\section{\label{secsix}Discussion}

We have addressed the question: should the number of fuzzball states equal $Exp[S_{bh}]$, the value suggested by black hole thermodynamics? It is difficult to explicitly construct and count all fuzzball states for a black hole. But we can use an indirect argument similar to the modular invariance relation used to count the states of a general 2-d CFT. We have followed the lines of \cite{hawkingisrael} where the Euclidean path integral $Z$ was computed in two ways: a saddle point evaluation and a Hamiltonian evolution. Implementing the changes required for the situation where all microstates are fuzzballs, we arrived at the relation (\ref{twtwo}). Once the entropy of fuzzballs agrees with the traditional entropy of the black hole, we argued that other thermodynamic quantities would have to agree as well.

In what follows,  we comment on the assumptions that have been made and  discuss the physical picture suggested by the `completeness' of fuzzball states. We end with some remarks on the Euclidean and Lorentzian sections of the black hole. 

\subsection{\label{secsix:1}The assumptions in the argument }

We have argued that $S_{fuzzball}=S_{bh}$, but in doing so we have made certain assumptions:

(i) The Euclidean path integral $Z$ in our theory of gravity has the same saddle point (\ref{fourt}) as was assumed in the Gibbons-Hawking computation.

(ii) The set of fuzzballs is complete; i.e., all states of the black hole are fuzzballs.

Since our argument is subject to these assumptions, we have not given a  `derivation' of the  agreement of entropies (\ref{twtwo}). But what we have argued is that under these qualitative assumptions,  the number of fuzzballs  would be a definite number - the  one given by the standard Bekenstein entropy - and not some other arbitrary number.  Note that these assumptions are a nontrivial statement about the physics of our gravity theory (string theory), since they can be violated for other theories:

(i')  Consider the 0+1 dimensional matrix model, which at low energies gives a 1+1 dimensional gravity theory. The low energy effective action has a Euclidean saddle point, but in the full theory quantum corrections invalidate the semiclassical approximation and prevent the formation of a  black hole \cite{polchinski}. Thus for this theory quantum fluctuations must invalidate the role of this saddle point as a leading order approximation to the path integral.

(ii') If we consider canonically quantized 3+1 gravity, then there are no fuzzball solutions. A similar situation holds for the 1+1 dimensional CGHS  \cite{cghs} and RST  \cite{rst} models.  

\subsection{\label{secsix:2}A physical picture of the path integral and its Hamiltonian decomposition}

Let us now give a physical picture of the computation of ${\cal N}_{fuzzball}$ given in section \ref{secfive}. We proceed in the following steps:

(a) It was argued in \cite{tunnel} that the semiclassical geometry of the Lorentizian black hole is destroyed by tunneling into the large number of possible fuzzball states. The fuzzball states are concentrated just outside $r=2M$, and so we expect the path integral in (\ref{seventq}) to be concentrated on metrics that live in the region $r\ge 2M$. We have assumed that this path integral has the saddle point (\ref{fourt}), and the $r, \tau$ space of this saddle point metric  is depicted in fig.\ref{c1}(a). We can picture the path integral in a cruude fashion  by imagining a lattice of points on this saddle point metric. Let these points be labelled by an index $\alpha$. At each lattice point $\alpha$, we have a variable $g_\alpha$ which symbolically represents the degrees of freedom of string theory at that point. The path integral $\prod_\alpha \int d g_\alpha e^{-S_{grav}[g]}$ gives $Z$.

\begin{figure}[h]
\includegraphics[scale=.38]{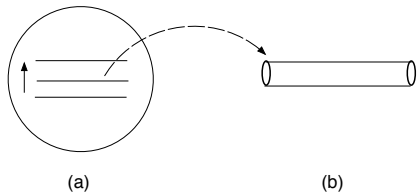}% Here is how to import EPS art
\caption{\label{c3} (a) A slicing of the $r, \tau$ space that is regular at the center of the disc  $r=2M$. (b) The slice through $r=2M$ will be  a cylinder, in the toy example where we assume there is one compact dimension.}
\end{figure}

\begin{figure}[h]
\includegraphics[scale=.42]{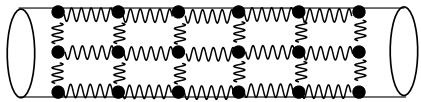}% Here is how to import EPS art
\caption{\label{c4} The field theory of the metric variables $g$ on the cylinder of fig.\ref{c3}(b). The spatial slice has been approximated by a lattice of points on which the variables $g_i$ live; the gradient term $\nabla g$ is represented by springs joining the lattice points.}
\end{figure}

(b) Before we move on to the Hamiltonian slicing leading to (\ref{eighttq}), consider the evaluation of $Z$ by a different slicing, indicated in fig.\ref{c3}(a). Here we have taken spacial slices that move smoothly through the central point $r=2M$. Recall that we have compact directions in our theory; these directions are important for the fuzzball construction. In fig.\ref{c3}(b)  we draw the slice passing through $r=2M$, with a  compact direction now explicitly exhibited as the circular direction of a cylinder.  In fig.\ref{c4} we depict the lattice points (from the above mentioned lattice) which lie on this cylinder; let these points be labelled by an index $i$. The variables $g_i$ at these lattice sites have kinetic terms $\nabla g$ which we have indicated by springs joining the sites.\footnote{We can get an approximate model of a  free quantum field theory by taking a set of masses joined by springs; we have used a crude model of this type to give a qualitative depiction of the point we wish to make.} The wavefunction on this slice is $\Psi[\{g_i\}]$, with $i$ running over  the lattice sites on the slice. Evolution along the slices of fig.\ref{c3}(a), with the appropriate Hamiltonian (which we call $\hat H'$) should reproduce $Z$.

\begin{figure}[h]
\includegraphics[scale=.42]{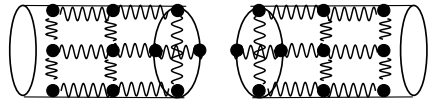}% Here is how to import EPS art
\caption{\label{c5} The same lattice of points as in fig.\ref{c4}, but with different springs connecting them. We can expand the same  state $\Psi[g_i]$ either as eigenfunctions of the Hamiltonian given by the links in fig.\ref{c4}, or by the links depicted above. In the case above, the inner ends of the cylinders have points identified with their diametrically opposite points, causing each of the cylinders to end in a cross-cap. }
\end{figure}

(c) Now consider the computation leading to (\ref{eighttq}). We take the same cylinder as in fig.\ref{c4}, with the same variables $g_i$ on the same lattice sites. But we imagine these sites to be joined by different links - the ones depicted in fig.\ref{c5}. The points on the cylinder have been divided into those on the left side and those on the right side.  At the end of each cylinder, a lattice site is linked to its diametrically opposite site. Thus the ends of the cylinders are `sewn-up', to produce a `capped' geometry.\footnote{In this simple example we get a `cross-cap', but in the full fuzzball geometry we expect the cap to be made of  KK-monopoles.} This set of `springs' on the links gives a different Hamiltonian $\hat H$ that acts on the {\it same} state $\Psi[\{g_i\}]$. Since the left and right sides of the cylinder are now disconnected, we can write the Hamiltonian as $\hat H=\hat H_L+\hat H_R$. Here $\hat H_L, \hat H_R$ are the Hamiltonians for the left and right sides, and act on states $\Psi_L, \Psi_R$ made out of the variables $g_i$ on the left and right sides respectively.\footnote{The difference between connected and disconnected slices in the Lorentzian section was discussed in the context of fuzzball complementarity in \cite{comp}. It was also discussed recently in \cite{ac} in the context of Maldacena's conjecture \cite{eternal} for  the eternal hole.}

We can expand the state   $\Psi_R$ in terms of eigenstates of the Hamiltonian $\hat H_R$. These eigenstates are  the analogues of the fuzzball states $|F_a\rangle$. We can get different eigenstates of $\hat H_R$  (with the same energy) by taking different excitations of the springs on the links. These different states correspond to the different possible `cap' states of the fuzzball. Taking a fuzzball state $|F_a\rangle$ and evolving by the Hamiltonian $\hat H_R$ around the $\tau$ circle in fig.\ref{c2} gives the contribution of this state to the path integral $Z$; let us call this contribution $Z_a$.\footnote{We do not have to separately consider the eigenstates of $\hat H_L$; as  $\tau$  sweeps around its full circle we automatically include the contribution of the left half of fig.\ref{c5}.}  Adding over the ${\cal N}_{fuzzball}$ values of the index $a$, we get the full partition function
\be
\sum_a Z_a=Z
\label{thirty}
\ee
We thus have a `fuzzball duality' which says that the contributions of all the fuzzballs add up to give the full path integral for the given compactification at infinity.  It is important to note that we should not take the path integral $Z$ in eq.(\ref{seventq}) and {\it also} take the contribution of fuzzball states; that would be overcounting. What happens instead is that the fuzzball states give a way of decomposing the full path integral into parts; each part arises from a state that has support only on manifolds that end without boundary in a `cap'.  It is a nontrivial statement that these fuzzball states are complete in the sense that they give (\ref{thirty}).  

\subsection{\label{secsix:3}Euclidean vs Lorentzian sections}

One may ask: what is the point of writing the full path integral $Z$ in terms of fuzzballs? If the goal is just to compute the Euclidean path integral $Z$, then we can imagine doing this path integral without the decomposition into fuzzball states.  But most of the interesting questions about black holes  lie in the Lorentzian section, where physical black hole processes can be considered. When we rotate back to the Lorentzian section $-i\tau\rightarrow t$, the eigenstates at energy $M$ are still the fuzzball states $|F_a\rangle$ at energy $M$. These states are wavefunctionals on the space of geometries that end compactly without boundary. Thus these states do not have the traditional horizon which would have the vacuum in its vicinity.\footnote{The absence of a traditional vacuum  is the definition of a fuzzball (see for example fig.6 in \cite{infofree}). The traditional horizon is `information-free', while the fuzzball state has the information of the state encoded in the cap structure at the horizon. Equivalently, one may consider the AMPS experiment \cite{amps} where an infalling observer seeks to measure the vacuum modes straddling the horizon; in a fuzzball state he would not find these modes in the state appropriate to the vacuum.} It is this fact which resolves the information paradox.

We should contrast the computation here with the approach of Sen \cite{sen}, where the goal is to count all the microstates of an extremal hole with a given charge. In the latter approach one finds the Bekenstien-Wald entropy \cite{wald} from the traditional geometry of the hole, and adds in contributions from `hair' modes. 
The result agrees perfectly with the count of states derived from D-brane counting along lines  similar to \cite{sv}. 

But as was noted in \cite{earlier}, the entropy obtained in this way from the  traditional horizon should not be taken as a count of states which individually have a traditional horizon. Rather, the traditional black hole metric used in \cite{sen} should be thought of being in the Euclidean section, i.e., the analogue of the  metric (\ref{fourt}). This is the case for two reasons:

(1) In the Lorentzian section, the traditional hole has a central singularity which can be accessed by a finite proper time of infall from the region outside the horizon. Thus, as a Lorentzian metric, the exterior of the horizon does not define a complete string background. If we extend the string background  to cover the interior of the hole, then we encounter the singularity which is not an allowed source in string theory. The approach of \cite{sen} requires an exact string background, so it is not clear how the traditional Lorentzian solution could be used.

(2)  The solution used in \cite{sen} is obtained by setting the coupling $g$ to zero in the duality frame where one is working. Thus this metric is a formal solution that is used as a tool to compute the entropy; it is not an actual state of the theory which would have to be defined at whatever (nonzero) value of the coupling we have. 

Thus the computation of \cite{sen} should be thought of as extending a Gibbons-Hawking type Euclidean computation in a way that yields the exact entropy  (for extremal holes) instead of just the leading order Bekenstein entropy. It should not be taken to imply that actual microstates have traditional horizons.

\begin{acknowledgments}
I thank Borun Chowdhury, Juan Maldacena, Ashoke Sen and David Turton for discussions.
This work was supported in part by DOE grant DE-FG02-91ER-40690.
\end{acknowledgments}

\nocite{*}

%\bibliography{counting}% Produces the bibliography via BibTeX.

\end{document}